\documentclass[twocolumn,showpacs,preprintnumbers]{revtex4-2}
\usepackage{amsfonts}
\usepackage{amssymb}
\usepackage{amsmath}
\usepackage{graphicx}
\usepackage{dcolumn}
\usepackage{bm}
\usepackage{color}
\usepackage{braket}
\usepackage[colorlinks=true, allcolors=blue]{hyperref}
\usepackage{float}
\usepackage{appendix}

\usepackage[utf8]{inputenc}

\begin{document}
	
	\title{Dissipative phase transitions of the  Dicke-Ising model}
    
    \author{Jun-Ling Wang$^{1}$}
	\author{Jiong Li$^2$}
	\author{Qing-Hu Chen$^{1,2,3,}$}
	\email{qhchen@zju.edu.cn}
	
	\affiliation{		
		$^1$ Zhejiang Key Laboratory of Micro-Nano Quantum Chips and Quantum Control, School of Physics, Zhejiang University, Hangzhou 310027, China. \\
		$^2$ Institute for Advanced Study in Physics, Zhejiang University, Hangzhou 310027, China \\ 
		$^3$ Collaborative Innovation Center of Advanced Microstructures, Nanjing University, Nanjing 210093, China.
	}	
	\date{\today}
	
	\begin{abstract}
        The dissipative phase transitions in the open transverse and longitudinal Dicke-Ising model (DIM), which incorporates nearest-neighbor Ising-type spin interactions into the Dicke framework, are investigated within a mean-field approach and further validated by detailed stability analysis. While the dissipative phase diagram of the transverse DIM is only slightly shifted upward compared with its ground‑state counterpart, dissipation in the longitudinal DIM stabilizes bistable nonequilibrium steady states and induces first-order phase transitions that are absent in the ground-state phase diagram. This bistable phase is characterized by the coexistence of superradiant and antiferromagnetic orders, and it converts a ground-state triple point into a tetracritical point, at which the boundaries of the first- and second-order transitions intersect. Our results reveal that the interplay among spin interactions, light-matter coupling, and dissipation supports a diverse set of nonequilibrium phase transitions and provides broad tunability of the phase diagram. These findings offer a theoretical foundation for exploring nonequilibrium physics in realistic open solid-state quantum systems.
	\end{abstract}

\maketitle

\section{Introduction}

The Dicke model, as a paradigmatic framework for investigating collective quantum phenomena, has been extensively studied over the past half-century due to its elegant theoretical structure and rich physical implications \cite{dicke_coherence_1954, hepp_superradiant_1973, wang_phase_1973}. It describes the collective linear coupling of $N$ two-level systems ("spins" or "qubits") to a single bosonic mode (e.g., a photon field). In the thermodynamic limit, the system undergoes a second-order quantum phase transition \cite{emary_quantum_2003, emary_chaos_2003, chen_numerically_2008, liu_large-_2009, zhuang2021universality,kirton_introduction_2019} from a normal phase to a superradiant phase characterized by macroscopic photon occupation. This transition has been experimentally reported in Bose-Einstein condensates within optical cavities \cite{nagy2010dicke,baumann_dicke_2010} and in simulations based on cavity-assisted Raman transitions \cite{dimer2007proposed, zhang_dicke-model_2018}, establishing the Dicke model as a cornerstone at the interface of quantum optics and condensed matter physics.

In its standard form, however, the Dicke model neglects direct interactions between material degrees of freedom, an assumption that is frequently violated in realistic physical systems. In solid-state platforms such as quantum dot arrays \cite{trif2007spin}, superconducting qubit networks \cite{pashkin2003quantum, tian2010circuit, johnson2011quantum}, and Rydberg atom ensembles \cite{duan2003controlling, nagy2010dicke}, unavoidable short- or long-range interactions—including dipole-dipole, exchange, or Coulomb interactions—naturally arise between neighboring spins. These intrinsic spin-spin interactions can substantially modify equilibrium and dynamical properties and may compete or cooperate with collective light-matter coupling, giving rise to quantum many-body phenomena beyond the pure Dicke model. 

Motivated by these considerations, the Dicke-Ising model (DIM) was proposed as a natural extension of the Dicke model \cite{lee_first-order_2004}. It incorporates Ising-type interactions on top of the Dicke Hamiltonian, thereby unifying photon-mediated long-range coherence (superradiant order) with interaction-driven local correlations (ferromagnetic or antiferromagnetic order). The equilibrium properties of this hybrid model, which lies at the interface of quantum optics and condensed matter physics, have recently garnered considerable research interest \cite{gammelmark_phase_2011, zhang_rydberg_2013, zhang_quantum_2014, luo_dynamic_2016, rohn_ising_2020, schuler_vacua_2020, schellenberger_almost_2024, schneider_dipolar_2024, langheld_quantum_2025, mendonca_role_2025, roman-roche_bound_2025, rao_unilateral_2025}. Various analytical and numerical many-body approaches have been applied to this system, including mean-field theories \cite{lee_first-order_2004, zhang_rydberg_2013, zhang_quantum_2014}, diagrammatic perturbation theory \cite{schneider_dipolar_2024}, exact diagonalization \cite{rohn_ising_2020, schuler_vacua_2020}, density matrix renormalization group \cite{mendonca_role_2025}, and wormhole quantum Monte Carlo techniques \cite{schneider_dipolar_2024}. Owing to the long-range nature of the light–matter interaction, mean-field predictions largely govern the overall phase behavior, while theories beyond mean field generally only make quantitative tuning to phase boundaries \cite{leibig_quantitative_2026}.

By contrast, the nonequilibrium dissipative phase transition (DPT) in the DIM has remained rarely explored. Although nonequilibrium DPTs in quantum many-body systems have become a central research topic at the interface of condensed matter physics and quantum optics, existing studies have primarily focused on the dissipative dynamics of the pure Dicke model \cite{keeling_collective_2010, torre_keldysh_2013, langford2017experimentally,doi:10.1073/pnas.1306993110, doi:10.1073/pnas.1417132112, kirton_introduction_2019}. This imbalance in theoretical understanding stands in sharp contrast to ongoing experimental progress. In particular, the well-known “no-go theorem” states that the $\mathrm{A}^2$ term arising from full electromagnetic interactions forbids equilibrium superradiant phase transitions \cite{rzazewski1975phase, nataf2010no, viehmann2011superradiant, ye2025superradiant}, thereby posing a fundamental obstacle to realizing equilibrium superradiant transitions in the pure Dicke model experimentally. To overcome this barrier, experimental approaches have shifted toward dissipation-driven nonequilibrium steady states, in which dissipation-induced superradiant phase transitions have been successfully observed in several solid-state platforms \cite{baumann_dicke_2010,doi:10.1073/pnas.1306993110,zhang_dicke-model_2018,kirton_introduction_2019}. 

Investigating DPTs in the DIM is of both fundamental theoretical interest and practical relevance. On the one hand, the inclusion of nearest-neighbor spin interactions renders this model a more faithful description of realistic platforms. On the other hand, dissipation is not only intrinsic to experimental setups but also serves as an additional “control knob” that competes with Ising interactions and light-matter coupling, enabling stable phases beyond equilibrium physics. Exploring phase transitions under the combined influence of dissipation and interactions is thus highly relevant for current solid-state quantum simulation platforms. In this work, we investigate how Ising interactions reshape the nonequilibrium steady-state phase diagram under dissipation and reveal novel nonequilibrium phases that arise from the interplay of dissipation, quantum coherence, and many-body interactions.

The paper is structured as follows. In Sec.~\ref{sec 2}, we introduce the transverse and longitudinal DIMs, discuss their underlying symmetries, and present and compare their ground-state phase diagrams.  Section~\ref{sec 3} derives the dissipative phase diagrams for both models. In particular, for the longitudinal DIM, we identify the emergence of a bistable phase accompanied by first-order dissipative phase transitions, underscoring genuinely nonequilibrium effects induced by dissipation. Finally, Section~\ref{sec.4} summarizes our main findings and discusses their physical implications.  Further details on the mean-field ground-state phase diagrams, steady-state solutions, and stability analysis are provided in the Appendices.

\section{Dicke-Ising Models and the ground-state phase diagrams}
\label{sec 2}	

The DIM is generally defined as the combination of a Dicke Hamiltonian and an Ising-type spin-spin interaction:
\begin{equation}
H_{\mathrm{DIM}}=H_{\mathrm{Dicke}}+H_{\mathrm{Ising}}.  \label{Hamiltonian1}
\end{equation}
The Dicke Hamiltonian takes the form \cite{emary_quantum_2003,kirton_introduction_2019}
\begin{equation}
H_{\mathrm{Dicke}} = \omega {a}^\dagger {a} +
\frac{\Omega}{2} \sum_{i=1}^N \sigma_{i}^{z} + \frac{g}{\sqrt{N}} \sum_{i=1}^N \left( {a}^\dagger + {a}\right)\sigma_{i}^{x},
\end{equation}
where $\Omega$ is the atomic transition frequency, and $\sigma _{i}^{\nu}$ ($\nu =x, y, z$) are the Pauli matrices of the $i$th spin, $a^{\dagger}(a)$ creates (annihilates) one photon in the common single-mode cavity with frequency $\omega$, $g$ denotes the dipole atom(spin)-cavity coupling strength, and $N$ is the total number of spins. 

To incorporate spin-spin interactions within the Dicke model framework, the Ising couplings can be formulated in two representative forms:
\begin{equation}
H_{\mathrm{Ising}}^{(z)} =J\sum_{\langle i,j\rangle }\sigma _{i}^{z}\sigma _{j}^{z},
\label{Ising_z}
\end{equation}
\begin{equation}
H_{\mathrm{Ising}}^{(x)}=J\sum_{\langle i,j\rangle }\sigma_{i}^{x}\sigma _{j}^{x}. 
\label{Ising_x}
\end{equation}
Here $J>0$ represents the antiferromagnetic coupling strength between nearest‑neighbor pairs $\langle i,j\rangle$, which can be implemented via various platforms, including circuit QED \cite{tian2010circuit}, optical cavity QED \cite{pashkin2003quantum, johnson2011quantum}, and Rydberg atomic systems \cite{duan2003controlling, nagy2010dicke}. The superscripts in $H_{\mathrm{Ising}}^{(z,x)}$ specify the orientation of the spin‑spin interaction in spin space. While the $z$‑directional variant has been extensively studied in the literature \cite{zhang_rydberg_2013, langheld_quantum_2025, mendonca_role_2025, shapiro2025digital}, we consider both possibilities. Depending on the relative orientation between the dipole qubit-cavity coupling axis ($\sigma_x$) and the Ising interaction axis, we refer to the Hamiltonian \eqref{Hamiltonian1} with $H_{\mathrm{Ising}}^{(z)}$ \eqref{Ising_z} as the \emph{transverse DIM}, and that with $H_{\mathrm{Ising}}^{(x)}$ \eqref{Ising_x} as the \emph{longitudinal DIM}.

To characterize the ordered phases, we introduce appropriate order parameters. For an Ising interaction oriented along the $\nu$‑direction with $J>0$, the antiferromagnetic order parameter is defined as the staggered magnetization
\begin{equation}
    m_{\mathrm{AF}}^\nu = \frac{1}{N} \sum_i (-1)^i \sigma_i^\nu.  \label{mdnu}
\end{equation}
In the superradiant phase, the simultaneous emergence of a finite collective dipole moment and a macroscopic photon population naturally identifies both as order parameters characterizing superradiant order. They are defined as:
\begin{equation}
   m_{\mathrm{DK}}^x = \frac{1}{N}\sum_i \sigma_i^x, \quad n = \langle a^\dagger a \rangle.  \label{mdx}
\end{equation}

\begin{figure}[tbp]
	\centering
	\includegraphics[width=1.0\linewidth]{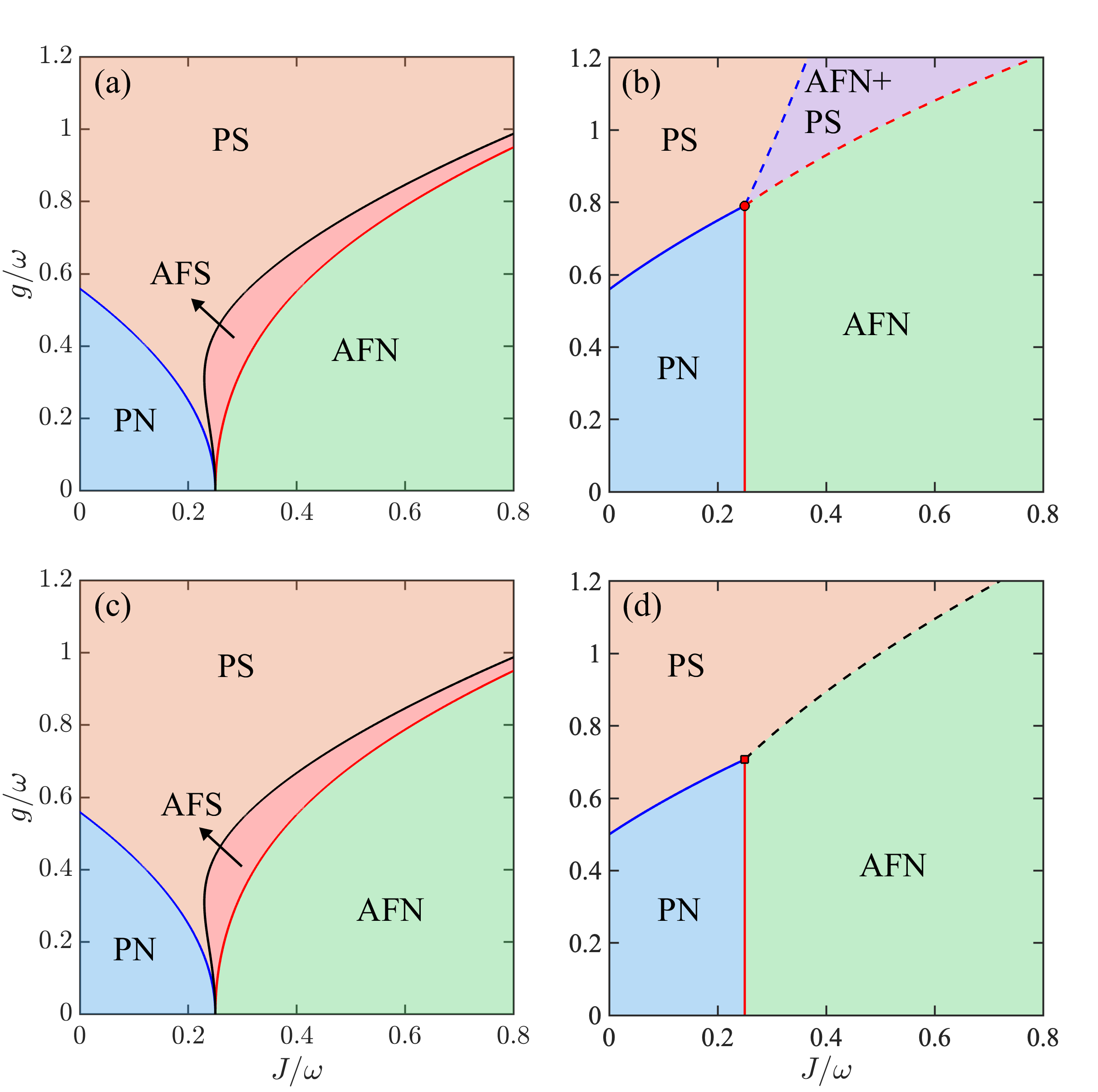}
	\caption{(a-b) Steady-state phase diagrams of the dissipative DIM in the $J$-$g$ plane with fixed $\Omega = \omega$ for the transverse DIM (a) and longitudinal DIM (b). The phases are labeled as PN (blue), AFN (green), PS (orange), AFS (red), and PS+AFN (purple). Solid blue, red, and black lines mark the boundaries of second-order transitions, while dashed blue and red lines indicate the first-order transition boundaries. The tetracritical point is highlighted by a red circle. The dissipative strength is fixed at $\kappa = 0.5\omega$. (c,d) Corresponding ground-state phase diagrams for (a,b). The black dashed line $g_{c0}^{x,\mathrm{GS}}$ indicates a first-order transition separating the AFN and PS phases. The red square represents the tricritical point.
}
\label{phase_diagram}
\end{figure}

The mean-field ground-state phase diagrams for both the transverse and longitudinal DIMs are shown in Fig.~\ref{phase_diagram} (c) and (d), respectively, with the corresponding mean-field ground-state analysis detailed in Appendix \ref{appendix-ground}. While the transverse DIM results have been reported previously in Ref.~\cite{zhang_quantum_2014} and are reproduced here for comparison, the longitudinal case shows a notable structural difference: an antiferromagnetic superradiant (AFS) phase, characterized by the coexistence of antiferromagnetic and superradiant orders, appears in the transverse DIM but is absent in the longitudinal DIM.

The qualitative differences between the two phase diagrams can be traced back to the underlying symmetries and competing interaction mechanisms. In the transverse DIM, when the atomic transition frequency vanishes ($\Omega = 0$), the spin-cavity coupling term $\sim\left( {a}^\dagger + a \right) \sigma_{i}^{x}$ plays the role of the longitudinal field in the Ising model, and the Hamiltonian possesses a $\mathbb{Z}_2$ symmetry associated with the spin-flip operation $\sigma_{i}^{z} \rightarrow -\sigma_{i}^{z}$. Similarly, in the longitudinal DIM, when the atom-cavity coupling vanishes ($g = 0$), the spin term $\sim \sigma^z$ acts as the transverse field in the Ising model, and the Hamiltonian also exhibits a $\mathbb{Z}_2$ symmetry, now realized through the spin-flip transformation $\sigma_{i}^{x} \rightarrow -\sigma_{i}^{x}$. For finite $\Omega$ and $g$, these spin-flip symmetries are explicitly broken. Nevertheless, for both the transverse and longitudinal DIMs, the full Hamiltonian \eqref{Hamiltonian1} commutes with the symmetry operator $P_d = - \hat{S}_z \exp \left(i\pi a^{\dagger }a\right)$ inherited from the pure Dicke model \cite{emary_chaos_2003}, where $\hat{S}_z \equiv \sum_{i} \sigma_{i}^z /2$ is the collective spin operator. Thus, the Dicke parity symmetry remains preserved in both DIMs.

In the transverse DIM, the Ising interaction is oriented perpendicular to the spin-cavity coupling direction, allowing antiferromagnetic and superradiant order parameters $m_{\mathrm{AF}}^z$ and $m_{\mathrm{DK}}^x$ to simultaneously become nonzero when the spins lie in the $x-z$-plane, thereby stabilizing an AFS phase. While such a phase is supported by quantitative approaches beyond mean-field theory \cite{langheld_quantum_2025, leibig_quantitative_2026}, mean-field treatments tend to overestimate its extent in the phase diagram.

By contrast, in the longitudinal DIM, both $m_{\mathrm{AF}}^x$ and $m_{\mathrm{DK}}^x$ are defined along the same $x$-direction. The short-range antiferromagnetic interaction favors staggered $\sigma_i^x$, whereas integrating out the photon field generates an effective long-range attractive interaction that promotes uniform spin alignment along the $\sigma^x$ direction. These two competing tendencies are fundamentally incompatible, thereby preventing the stabilization of an AFS phase and explaining its absence in Fig.~\ref{phase_diagram} (d). In the regime of weak spin‑cavity coupling $g$, the system is effectively governed by a standard transverse Ising model $\frac{\Omega}{2} \sum_{i=1}^N \sigma_{i}^{z}+ J\sum_{\langle i,j\rangle }\sigma_{i}^{x}\sigma _{j}^{x}$, so that a small antiferromagnetic interaction $J$ is insufficient to induce staggered spin order, leaving the system in the paramagnetic normal (PN) phase. Once $J$ exceeds a critical value, an antiferromagnetic normal (AFN) phase appears. In the strong‑coupling regime, the excitation of a macroscopic photon population, combined with dominant long‑range spin attraction, drives the system into the paramagnetic superradiant (PS) phase.

The interplay of $\mathbb{Z}_2$ antiferromagnetic and $\mathbb{Z}_2$ Dicke parity symmetries allows, in principle, for a maximum of four distinct phases. Indeed, all four phases are realized in the transverse DIM. In the longitudinal DIM, however, only three phases appear, with the AFS phase absent. These ground-state results provide a key reference for the nonequilibrium dissipative scenario discussed in the next section. Dissipation fundamentally modifies the mechanisms that govern phase selection, enabling phases absent in the ground state to become stable nonequilibrium steady states. This leads to a central question: How does dissipation reshape the phase structure of the DIMs?

\section{Dissipative phase diagram}

\label{sec 3}

We now turn to the nonequilibrium behavior of the DIM in the presence of dissipation. Instead of relaxing into the lowest-energy ground state, the system evolves toward a nonequilibrium steady state governed by the dynamical competition between coherent Hamiltonian dynamics and dissipative processes, which introduce damping and decoherence.

In an open DIM, the cavity frequency typically constitutes the largest energy scale in the system. As a result, the reservoir can be well approximated as Markovian and generally violates the equilibrium fluctuation-dissipation relation. Under the rotating-wave and the Born-Markov approximations, the dissipative dynamics of the system can be described by a Lindblad master equation in a local time frame
\begin{equation}
	\dot{\rho} = -i\left[H_{\mathrm{DIM}},\rho\right] + \kappa \mathcal{D}\left[a\right],
	\label{lindblad}
\end{equation}
where $\mathcal{D}[a] = 2 a \rho a^\dagger - a^\dagger a \rho - \rho a^\dagger a$ is the dissipative superoperator associated with the cavity damping rate $\kappa$. 

Importantly, the cavity damping term in Eq.~\eqref{lindblad} does not break the $\mathbb{Z}_{2}$ symmetry, as there is no explicit driving field that would violate the symmetry. Consequently, the open DIM preserves its $\mathbb{Z}_{2}$-symmetry, thereby allowing for the emergence of dissipative steady-state phase transitions.

\subsection{Dissipative transverse DIM}

We first analyze the transverse DIM under dissipation. In this configuration, the Ising interaction is oriented perpendicular to the spin–cavity coupling axis. The mean-field nonequilibrium steady-state solutions are obtained by solving the corresponding semiclassical equations of motion. Specifically, by neglecting quantum fluctuations and factorizing operator expectation values, the mean-field equations can be derived from Eq.~\eqref{Hamiltonian1} within the Lindblad formalism of Eq.~\eqref{lindblad}, yielding
\begin{equation}
	\begin{aligned}
		{\braket{\dot{\sigma}_{A(B)}^x }}&= -\left(\Omega+4J\braket{\sigma_{B(A)}^z}\right)\braket{\sigma_{A(B)}^y}, \\
		{\braket{\dot{\sigma}_{A(B)}^y}} &= \left(\Omega+4J\braket{\sigma_{B(A)}^z}\right)\braket{\sigma_{A(B)}^x} \\
        &-\frac{4g}{\sqrt{N}}\mathrm{Re}\braket{a}\braket{\sigma_{A(B)}^z},\\
		{\braket{\dot{\sigma}_{A(B)}^z }}&= \frac{4g}{\sqrt{N}}\mathrm{Re}\braket{a}\braket{\sigma_{A(B)}^y},\\
        \dot{\braket{a}} &= - i\left(\omega - i\kappa\right)\braket{a} - i\frac{g}{\sqrt{N}}\sum_i \braket{\sigma_i^x},
	\end{aligned}
\end{equation}
where the antiferromagnetic interaction naturally partitions the lattice into two sublattices, labeled $A$ and $B$. 

For convenience, we introduce the rescaled parameters $\alpha = \langle a \rangle / \sqrt{N}$ and $s_\mu^\nu = \langle \sigma_\mu^\nu \rangle$ ($\mu = A, B$). The mean displacement $\alpha$ is further decomposed into real and imaginary parts, $\alpha = \alpha_R + i \alpha_I$. In terms of these variables, the steady-state conditions are given by 
\begin{equation}
	\begin{aligned}
		0 &= -\kappa \alpha_R + \omega \alpha_I, \\
		0 &= -\kappa \alpha_I - \omega \alpha_R - \frac{g}{2}\left(s_A^x + s_B^x\right), \\
		0 &= \left(\Omega + 4Js_{B(A)}^z\right)s_{A(B)}^x-4g\alpha_R s_{A(B)}^z.
	\end{aligned}
	\label{steady-state_z}
\end{equation}
Additionally, the steady-state solutions satisfy $s_\mu^y=0$ for each sublattice, and spin-length conservation imposes the constraint $s_\mu^{x2} + s_\mu^{z2} = 1$. 

Owing to the symmetry of the system, the steady-state solutions of the transverse DIM can be classified into four distinct phases, as shown in Fig.~\ref{phase_diagram} (a) [see Appendix \ref{appendix steady-state solution} for more details]. Using the notation  $\ket{\searrow}$, $\ket{\swarrow}$, $\ket{\nearrow}$, and $\ket{\nwarrow}$ for atomic spin states polarized in the $x$(horizontal)-$z$(vertical) plane, and  $\ket{\uparrow}$, $\ket{ \downarrow }$ for spin-up and spin-down state along the $z$-direction, we characterize these four phases as follows.

\textbf{(i)} \emph{Paramagnetic normal (PN) phase-:} The system has a unique steady state: $\ket{\cdots \downarrow \downarrow \downarrow \downarrow \cdots} \otimes \ket{0}$, characterized by $n=m_{\mathrm{DK}}^{x}=0$ and $m_{\mathrm{AF}}^{z}=0$.

\textbf{(ii)} \emph{Antiferromagnetic normal (AFN) phase-:}  Two degenerate steady states exist: $\ket{\cdots \downarrow \uparrow \downarrow \uparrow \cdots} \otimes \ket{0}$ and $\ket{\cdots \uparrow \downarrow \uparrow \downarrow \cdots}\otimes \ket{0}$, characterized by  $n=m_{\mathrm{DK}}^x=0$ and $m_{\mathrm{AF}}^{z}=1$.

\textbf{(iii)} \emph{Paramagnetic superradiant (PS) phase-:} Two degenerate steady states are: $\ket{\cdots \swarrow \swarrow \swarrow \swarrow \cdots}\otimes \ket{\alpha}$ and $\ket{\cdots \searrow \searrow \searrow \searrow \cdots}\otimes \ket{-\alpha}$, characterized by $n\neq 0$ and $m_{\mathrm{DK}}^x\neq 0$. Here the nonzero antiferromagnetic  order parameters satisfy  $m_{\mathrm{AF}}^{z}=g^{2}\left[ 1-\Omega^{2} / (16J_{\mathrm{eff}}) \right]/( \omega^{2}+\kappa^{2})$, where $J_{\mathrm{eff}}=J+\omega g^{2}/\left( \omega ^{2}+\kappa ^{2}\right)$. 

\textbf{(iv)} \emph{AFS phase-:}  Two degenerate steady states are: $\ket{\cdots \nearrow \searrow \nearrow \searrow \cdots} \otimes \ket{\alpha}$ and $\ket{\cdots \searrow \nearrow \searrow \nearrow \cdots}\otimes \ket{-\alpha}$, characterized by $n\neq 0, m_{\mathrm{DK}}^x \neq 0$ and $m_{\mathrm{AF}}^{z} \neq 0$. The explicit expressions of these order parameters are provided in Appendix~\ref{appendix steady-state solution}. 

To construct a complete dissipative phase diagram, we examine the dynamical stability of the mean-field solutions by considering small fluctuations around the mean-field steady states: $\alpha \rightarrow \alpha + \delta\alpha$, and $s_\mu^\nu \rightarrow s_\mu^\nu + \delta s_\mu^\nu$. Retaining only terms linear in these fluctuations yields a set of linearized equations of motion around the mean-field solutions, which govern the dynamics of small perturbations and thus determine their dynamical stability. The fluctuation vector is defined as $\delta \mathbf{X} = (\delta\alpha_R, \delta\alpha_I, \delta s_A^x, \delta s_A^y, \delta s_A^z, \delta s_B^x, \delta s_B^y, \delta s_B^z)^\top$, whose time evolution obeys $\delta \dot{\mathbf{X}} = \mathbf{M} \delta \mathbf{X}$, with $\mathbf{M}$ denoting stability matrix. The stability conditions for all steady-state phases are obtained from the eigenvalue spectrum of $\mathbf{M}$ [see Appendix \ref{appendix stability analysis} for more details].

\textbf{(i)} \emph{PN phase-:} $J < J_c = \Omega/4$ and $g < g_{c1}^z = \sqrt{\left( \Omega - 4J \right)(\omega ^{2}+\kappa ^{2})/ \left(4\omega\right)}$;

\textbf{(ii)} \emph{AFN phase-:} $J>J_c$ and $g < g_{c2}^z = \sqrt{\left( 16J^2 - \Omega^2 \right)(\omega ^{2}+\kappa ^{2}) / \left( 16\omega J \right)}$;

\textbf{(iii)} \emph{PS phase-:} $g > g_{c1}^z$ and $16J_{\mathrm{eff}} - 32JJ_{\mathrm{eff}} + J \Omega^2 > 0$; 

\textbf{(iv)} \emph{AFS phase-:} $g>g_{c2}^z$ and $16 J_{\mathrm{eff}} - 32JJ_{\mathrm{eff}} + J \Omega^2 < 0$.

For the transverse DIM, the overall structure of the dissipative phase diagram remains qualitatively similar to that of the ground-state phase diagram, with the phase boundaries shifted slightly toward larger coupling strengths. This behavior closely parallels that of the pure Dicke model, in which the critical coupling $g_c$ for the dissipative phase transition is enhanced by a scaling factor $\sqrt{(\omega ^{2}+\kappa ^{2})/{\omega^2 }}$ \cite{kirton_introduction_2019}. 

\begin{figure}[tbp]
	\centering
	\includegraphics[width= 0.9\linewidth]{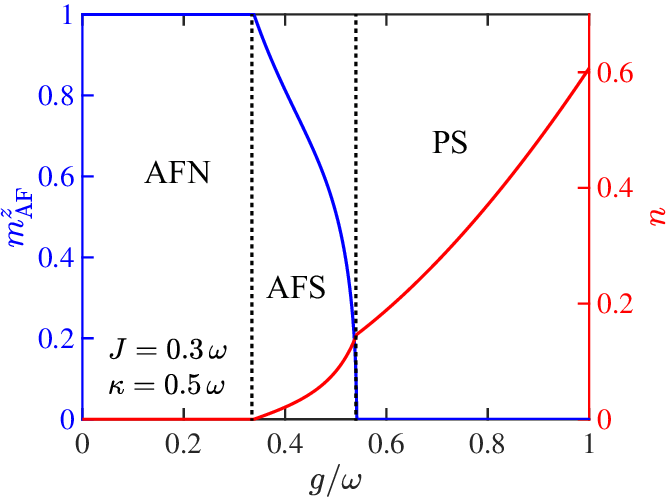}
    \caption{Order parameters $m_{\mathrm{AF}}^z$ and $n$ as functions of the coupling strength $g$ in the dissipative transverse DIM with $J = 0.3\omega$. The blue and red solid lines correspond to $m_{\mathrm{AF}}^z$ and $n$, respectively. The dissipation rate is fixed at $\kappa = 0.5\omega$.
    }
    \label{order_z_diss}
\end{figure}

In Fig.~\ref{order_z_diss}, we present both the antiferromagnetic order parameter $m_{\mathrm{AF}}^z$ and the average photon number $ n $ as functions of the coupling strength $g$ at fixed $J=0.3 \omega$ in the steady state. Both $m_{\mathrm{AF}}^z$ and $n$ evolve continuously across the critical points, while their first-order derivatives clearly exhibit discontinuities, confirming that the associated phase transitions are of second order.

Since all four ground-state phases of the transverse DIM remain stable in its dissipative counterpart, a key question emerges for the longitudinal case: how does dissipation shape the phase behavior of the longitudinal DIM? Notably, in this model, the AFS phase with both antiferromagnetic and superradiant orders is absent in the ground state. We address this question in the following subsection.

\subsection{Dissipative longitudinal DIM}

We now turn to the longitudinal DIM, in which antiferromagnetic and superradiant orders directly compete for the same spin component. Here too, we focus on the steady-state solutions and their dynamical stability.

Following the same mean-field procedure as in the previous subsection, the steady-state conditions are obtained from the semiclassical equations of motion within the Lindblad formalism:
\begin{equation}
	\begin{aligned}
		0 &= -\kappa \alpha_R + \omega \alpha_I, \\
		0 &= -\kappa \alpha_I - \omega \alpha_R - \frac{g}{2}\left(s_A^x + s_B^x\right), \\
		0 &= \Omega s_{A(B)}^x - 4\left(g \alpha_R + J s_{B(A)}^x\right) s_{A(B)}^z.
	\end{aligned}
	\label{steady-state_x}
\end{equation}

The steady-state solutions of the longitudinal DIM can be classified into three distinct phases, as shown in Fig.~\ref{phase_diagram} (b) [see Appendix~\ref{appendix steady-state solution} for more details]. Using the same  notation for the atomic spin states in the $x$(horizontal)-$z$vertical) plane as those in the previous subsection, we characterize these three phases as follows.

\textbf{(i)} \emph{PN phase-:}  A single steady state, $\ket{\cdots \downarrow \downarrow \downarrow \downarrow \cdots} \otimes \ket{0}$, characterized by $n
= m_{\mathrm{DK}}^x = 0$ and $m_{\mathrm{AF}}^x = 0$.

\textbf{(ii)} \emph{AFN phase-:} Two degenerate steady states, $\ket{\cdots \swarrow \searrow
\swarrow \searrow \cdots} \otimes \ket{0}$ and $\ket{\cdots \searrow \swarrow \searrow \swarrow \cdots} \otimes \ket{0}$, characterized by $n = m_{\mathrm{DK}}^x = 0$ and $m_{\mathrm{AF}}^x= \sqrt{1 - {\Omega^2}/{16J^2}}$.

\textbf{(iii)} \emph{PS phase-:} Two degenerate steady states are: $
\ket{\cdots \swarrow \swarrow \swarrow \swarrow \cdots}\otimes\ket{\alpha}$ and $\ket{\cdots \searrow \searrow \searrow \searrow \cdots}\otimes\ket{-\alpha}$, characterized by $n = g^2 \left[1-(\Omega/4J_{\mathrm{eff}})^2 \right]/\left(\omega^2+\kappa^2\right)$, $m_{\mathrm{DK}}^x=\sqrt{1-({\Omega}/{4J_{\mathrm{eff}})}^2}$, and $m_{\mathrm{AF}}^x = 0$, where $J_{\mathrm{eff}} = J - \omega g^2/(\omega^2 + \kappa^2)$ denotes the effective Ising interaction.

Strikingly, the AFS phase is absent in the longitudinal DIM, in close analogy with the corresponding ground-state phase diagram. The complete dissipative phase diagram is obtained through a stability analysis following the same procedure as for the transverse DIM. Then we obtain the stability condition for all phases [see Appendix \ref{appendix stability analysis} for more details]:

\textbf{(i)} \emph{PN phase-:} $J<J_c = \Omega/4$ and $g<g_{c1}^x = \sqrt{\left(\Omega + 4J\right) (\omega ^{2}+\kappa ^{2}) / \left( 4\omega \right)}$;

\textbf{(ii)} \emph{AFN phase-:} $J>J_c$ and $g < g_{c2}^x = \sqrt{J \left( 16J^2 + \Omega^2 \right)(\omega ^{2}+\kappa ^{2})/\left( \omega \Omega^2 \right)}$;

\textbf{(iii)} \emph{PS phase-:} $g>g_{c1}^x$ for $J<J_c$ and $g>g_{c3}^x = \sqrt{\left[ J + \left( J\Omega^2 / 16 \right)^{1/3} \right] (\omega ^{2}+\kappa ^{2})/ \omega}$ for $J>J_c$.

\textbf{(iv)} \emph{Bistable phase (ANF+PS)-:} $g_{c3}^{x}<g<g_{c2}^x$ for $J>J_c$.

The steady-state phase diagram of the longitudinal DIM is shown in the $J$–$g$ plane for a fixed $\Omega = \omega$, as presented in Fig.~\ref{phase_diagram}(b). In the weak coupling regime, increasing $J$ leads to a second-order DPT from the PN phase to the AFN phase. Similarly, for $J < J_c$, increasing $g$ induces a second-order transition from the PN phase to the PS phase. The dashed blue and red lines indicate the loss of dynamical stability of the AFN and PS phases, respectively. This gives rise to a \emph{bistable phase} where the AFN and PS steady states coexist in the region bounded by $g_{c3}^x<g<g_{c2}^x$. Notably, the three critical couplings $g_{c1}^x$, $g_{c2}^x$, and $g_{c3}^x$ merge at $J = J_c$, forming a tetracritical point 
\begin{equation}
    g_{\mathrm{tet}} = \sqrt{\Omega\left(\omega^2 + \kappa^2\right)/\left(2 \omega\right)},
\end{equation} 
which is marked by a red circle in Fig.~\ref{phase_diagram} (b).

Specifically, when $J = 0$, the system reduces to the pure Dicke model and undergoes a second-order DPT from normal phase to superradiant phase when $g$ exceeds the critical coupling constant of the dissipative pure Dicke model $\sqrt{\Omega\left(\omega^2 + \kappa^2\right)/\left(4\omega\right)}$ \cite{kirton_introduction_2019}. In the absence of spin-cavity coupling ($g = 0$), the system recovers the mean‑field ground‑state phase transition of the transverse Ising model with critical coupling $J_c=\Omega/4$, as atomic decay is not taken into account here. Finally, in the dissipationless limit ($\kappa = 0$), $g_{c1}^x$ reduces to $g_{c1}^{x,\mathrm{GS}} = \sqrt{\omega\left(J + \Omega/4\right)}$, which corresponds to the second-order phase boundary between the PN and the PS phases in the ground state.

\begin{figure}[tbp]
	\centering
	\includegraphics[width= 0.9\linewidth]{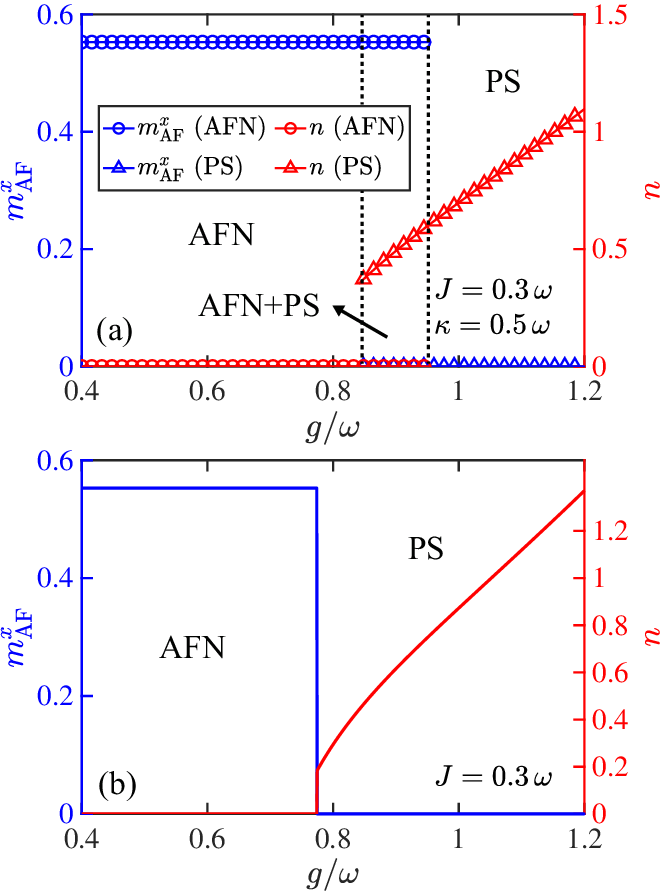}
    \caption{Order parameters as functions of the coupling strength $g$ at $J = 0.3\omega$ for open longitudinal DIM with $\kappa = 0.5\omega$ (a), and the corresponding closed system with $\kappa = 0$ (b). Blue lines represent the staggered magnetization along the $x$ direction, $m_{\mathrm{AF}}^x$, while red lines represent the mean photon number $n$. Circular markers correspond to values obtained in the AFN phase, whereas triangular markers indicate results in the PS phase.
    }
    \label{order_x}
\end{figure}

Dissipative bistability in the longitudinal DIM describes a dynamical regime in which the system admits two possible steady states in its phase diagram. Each phase is characterized by a single dominant order parameter—either the antiferromagnetic order ($m_{\mathrm{AF}}^x$) or the photon number ($n$). Which of the two states is realized depends predominantly on the initial conditions. Dissipation plays a crucial role in enabling this form of “coexistence.” In the closed limit ($\kappa\rightarrow 0$), the system selects the lower-energy phase unambiguously, and bistability is absent. By contrast, in the presence of dissipation, the dependence on initial conditions effectively opens a finite parameter region in which either order can be stabilized as a steady state. 

The nature of the DPTs is further clarified through the evolution of the order parameters for the longitudinal DIM, presented in Fig.~\ref{order_x}(a) for $J = 0.3\omega$. As $g$ increases, the system passes through a sequence of DPTs, evolving from the AFN phase, through a region of bistable AFN+PS coexistence, to the pure PS phase. If the PS branch is selected upon entering the bistable phase, the photon number $\langle a^\dagger a \rangle$ jumps from zero to a finite value, while the staggered magnetization drops to zero, indicating a first-order transition. By contrast, if the AFN branch is selected, $m_{\mathrm{AF}}^x$ remains finite and $\langle a^\dagger a \rangle$ stays zero as the system evolves from the AFN phase into the bistable regime. Upon further increasing $g$ beyond $g_{c2}^x \approx 0.957\omega$, the AFN branch loses dynamical stability, and the system enters the PS phase through another first-order transition. Thus, for a fixed $J > J_c$, the system exhibits successive first-order phase transitions as a function of $g$: AFN $\rightarrow$ AFN+PS bistable $\rightarrow$ PS phases.

In Fig.~\ref{order_x} (b), the corresponding order parameters $m_{\mathrm{AF}}^x$ and  $\langle a^\dagger a \rangle$ are plotted as functions of $g$ for the closed longitudinal DIM. Both quantities exhibit discontinuous jumps, signaling a direct first-order phase transition from the AFN to the PS phase. Thus, for the longitudinal DIM, the dissipative phase diagram differs markedly from its ground-state counterpart: dissipation stabilizes a bistable region within the PS sector at larger values of $J$. Meanwhile, the remaining phase boundaries also shift upward slightly, a trend consistent with that observed in the pure Dicke model.

The behavior of the order parameters in the bistable phase differs fundamentally from that in the AFS phase of the closed transverse DIM. As illustrated in Fig.~\ref{order_z_diss}, the antiferromagnetic order parameter varies continuously between $0$ and $1$ in the AFS phase, reflecting the simultaneous coexistence of antiferromagnetic and superradiant orders. In contrast, in the bistable phase of the dissipative longitudinal DIM, the same order parameter only takes the discrete values $1$ or $0$, corresponding respectively to the pure  AFN and PS steady states.

Finally, we address the order of the phase transitions within a mean-field Landau framework. According to the Landau theory, a phase transition between two phases characterized by different broken symmetries is typically of first order. This criterion applies directly to the transition between the AFN and PS phases: the AFN phase breaks translational (antiferromagnetic) symmetry, whereas the PS phase breaks the $\mathbb{Z}_2$ symmetry associated with superradiant order. Consequently, the AFN-PS transition is first order, which is consistent with the discontinuous jumps of the order parameters observed in the longitudinal DIM.

In the transverse DIM, by contrast, an intermediate AFS phase emerges between the PS and AFN phases. Importantly, both the PS and AFN phases already break a $\mathbb{Z}_2$ symmetry: the PS phase breaks the Dicke-type $\mathbb{Z}_2$ symmetry associated with superradiant order, while the AFN phase breaks the Ising-type $\mathbb{Z}_2$ symmetry associated with antiferromagnetic order. Upon entering the AFS phase from the PS side, the Ising-type $\mathbb{Z}_2$ symmetry is broken in addition to the already broken Dicke-type $\mathbb{Z}_2$ symmetry. Conversely, the transition from AFN to AFS involves the onset of Dicke-type $\mathbb{Z}_2$ symmetry breaking on top of the preexisting Ising-type symmetry breaking. In both cases, the transition involves the emergence of an additional order parameter rather than a switch between phases with incompatible broken symmetries, and therefore, both the PS-AFS and AFN-AFS transitions are continuous and of second order.

These results establish that dissipation does not merely shift phase boundaries in the longitudinal DIM but qualitatively reshapes the phase diagram by stabilizing bistable nonequilibrium steady states and inducing first-order phase transitions. The sharp contrast between the longitudinal and transverse DIMs highlights the crucial role played by the relative orientation between spin-spin interactions and light–matter coupling.

\section{\label{conclusion}conclusion}
\label{sec.4}

In this work, we have systematically investigated DPT in the DIMs, in which nearest-neighbor Ising-type spin interactions are incorporated into the Dicke framework, using a mean-field approach combined with stability analysis. By focusing on two distinct realizations--the transverse and longitudinal DIMs--we have elucidated how the interplay between spin-spin interactions, collective light-matter coupling, and dissipation qualitatively reshapes the nonequilibrium steady-state phase structure beyond equilibrium physics.

For the transverse DIM, where the Ising interaction is oriented perpendicular to the spin-cavity coupling, we find that the overall structure of the steady-state phase diagram remains qualitatively similar to that of the ground state. An intermediate AFS phase persists in the presence of dissipation, separating the PS and AFN phases. The associated phase transitions are continuous and of second order, reflecting a sequential symmetry-breaking mechanism whereby the Dicke-type and Ising-type $\mathbb{Z}_2$ symmetries are broken successively. In this case, dissipation primarily leads to a quantitative renormalization of the phase boundaries, closely analogous to the behavior observed in the pure Dicke model.

By contrast, the longitudinal DIM exhibits qualitatively new nonequilibrium behavior. Because the Ising interaction and the spin-cavity coupling act along the same spin direction, antiferromagnetic and superradiant orders compete directly. As a result, the AFS phase is absent, and dissipation stabilizes a bistable steady-state region in which AFN and PS phases coexist. This bistability gives rise to first-order DPTs, accompanied by discontinuous jumps in the order parameters. Such behavior has no counterpart in the ground-state phase diagram and therefore represents a genuine nonequilibrium effect induced by dissipation. From a symmetry perspective, the direct transition between phases characterized by incompatible broken symmetries naturally accounts for the emergence of first-order transitions in the longitudinal DIM, in sharp contrast to the continuous transitions found in the transverse case.

More generally, our results highlight that dissipation does not merely shift equilibrium phase boundaries, but can qualitatively reshape the phase structure of interacting light-matter systems by stabilizing new steady states and inducing bistability. While our analysis is based on a mean-field treatment, the symmetry-based arguments and the qualitative distinction between transverse and longitudinal DIMs suggest that the emergence of bistability and first-order transitions is robust beyond mean field. The DIM thus provides a versatile platform for exploring nonequilibrium many-body physics arising from the interplay of competing orders and dissipation. In current solid-state platforms where strongly coupled superradiant phase transitions can be realized, spin-spin interactions are generically present and cannot be neglected, while environmental dissipation is intrinsic and unavoidable.
     
Finally, we note that the Ising interaction may also be oriented along the $y$ direction, as considered in Refs.\cite{lee_first-order_2004,rao_unilateral_2025}, leading to a fully noncommuting spin algebra among the three spin components, which is expected to support an even richer and more intrinsically quantum phase diagram. Moreover, extending the spin-spin coupling to more complex forms, such as Heisenberg-type \cite{mendonca_role_2025}, XY-type \cite{al-saidi_several_2002, chen_quantum_2010}, or long-range interactions \cite{koziol2025melting}, would be highly valuable for future studies aimed at more faithfully describing realistic experimental platforms and quantum simulation settings. In this sense, the present work establishes a solid foundation for future theoretical and experimental investigations of nonequilibrium phase transitions in open quantum many-body systems.

\begin{acknowledgments}
This work is supported by the National Key R\&D Program of China  (Grants No. 2024YFA1408900 and No. 2022YFA1402701), the National Natural Science Foundation of China under Grant No. 92565201.
\end{acknowledgments}

\appendix

\section{\label{appendix-ground} GROUND-STATE PHASE TRANSITIONS OF THE TRANSVERSE DIM AND LONGITUDINAL DIM}

In this Appendix, we provide detailed derivations of the mean-field ground-state solutions for both the transverse and longitudinal DIM models. Although the transverse model has already been analyzed in Ref.\cite{zhang_quantum_2014}, we present its derivation here in our own notation and framework, both for completeness and to facilitate direct comparison with the longitudinal case.

\emph{Transverse DIM}-:
Considering the Hamiltonian \eqref{Hamiltonian1} for the transverse DIM in the absence of dissipation, we adopt the mean-field approach $\alpha = \langle a \rangle / \sqrt{N}$ and $s_\mu^\nu = \langle \sigma_\mu^\nu \rangle$ ($\mu = A, B$) to obtain the scaled  ground-state energy as
\begin{equation}
    \begin{aligned}
        E_g &= \omega \alpha^2 + g\alpha\left(s_A^x + s_B^x\right) + J s_A^z s_B^z + \frac{\Omega}{4}\left(s_A^z +s_B^z\right).
	\label{ground_energy_x}
    \end{aligned}
\end{equation}
By minimizing the ground-state energy with respect to $\alpha$, we obtain:
\begin{equation}
	 \frac{\partial E_g}{\partial \alpha} = 2\omega\alpha + g\left(s_A^x + s_B^x\right).
\end{equation}
Then we obtain the effective ground-state energy for spin by substituting $\alpha = -g\left(s_A^x + s_B^x\right)/(2\omega)$ into \eqref{ground_energy_x}
\begin{equation}
    \begin{aligned}
        E_g &= -\frac{g^2}{4\omega}\left(s_A^x + s_B^x\right)^2 + J s_A^z s_B^z + \frac{\Omega}{4}\left(s_A^z + s_B^z\right).
    \end{aligned}
\end{equation}
Due to the antiferromagnetic interaction, we parameterize the spin components as $s_{A(B)}^x = \cos \theta_{1(2)}$ and $s_{A(B)}^z=\pm \sin\theta_{1(2)}$, which leads to rewriting the ground-state energy:
\begin{equation}
    \begin{aligned}
        E_g &= -\frac{g^2}{\omega}\left(\cos \theta_1 + \cos \theta_2\right)^2 - J\sin \theta_1 \sin\theta_2 \\
        &+ \frac{\Omega}{4}\left(\cos \theta_1 + \cos \theta_2\right) 
    \end{aligned}\label{Eg_x_theta}
\end{equation}

To further simplify the analysis, we introduce the symmetric and antisymmetric combinations $a=(\theta_1 + \theta_2)/2$ and $b=(\theta_1-\theta_2)/2$, respectively. Under this definition, the order parameters becomes $m_{\mathrm{AF}}^x=|\sin a \sin b|$, $m_{\mathrm{AF}}^z =|\sin a \cos b|$ and $n=g^2\cos^2 a \cos^2b/\omega^2$. Then Eq.~\eqref{Eg_x_theta} is rewritten as
\begin{equation}
    \begin{aligned}
        E_g &= \frac{\Omega}{2}\cos a \sin b -\frac{g^2}{\omega}\cos^2a\cos^2b-J\left(\sin^2a -\sin^2b\right)     
    \end{aligned}
\end{equation}
Thus, we take partial derivatives with respect to $a$ and $b$, leading to the stationary point equations
\begin{equation}
    \begin{aligned}
        \sin a\left[2\left(\frac{g^2}{\omega}\cos^2 b -J\right)\cos a - \frac{\Omega}{2}\sin b\right] &=0,\\
        \cos b\left[2\left(\frac{g^2}{\omega}\cos^2 a +J\right)\sin b +\frac{\Omega}{2}\cos a\right] &= 0.
    \end{aligned}
\end{equation}
The solutions can be classified into four distinct cases.

\textbf{(i)} PN phase ($\sin a = \cos b =0$): .
\begin{equation}
    s_A^x =s_B^x = 0,\;s_A^z=s_B^z=-1,\;\alpha=0.
\end{equation}

\textbf{(ii)} AFN phase ($\sin b = \cos a=0$): 
\begin{equation}
    s_A^x = s_B^x=0,\;s_A^z=-s_B^z= \pm1,\;\alpha=0.
\end{equation}

\textbf{(iii)} PS phase ($\sin a=0$ and $\sin b \neq0$ ): 
\begin{equation}
    \begin{aligned}
        s_A^x&=s_B^x=\pm\sqrt{1-s_{A(B)}^{z2}},\;s_A^z = s_B^z = - \frac{\Omega}{4J_{\mathrm{eff}}},\\
        \alpha&=-\frac{g}{\omega}s_{A(B)}^x,
    \end{aligned}
\end{equation}
where $J_{\mathrm{eff}} = J+ g^2/\omega$. 

\textbf{(iv)} AFS phase ($\sin a \neq0$ and $\cos b \neq0$): 
\begin{equation}
    \begin{aligned}
        \cos a &= \frac{\Omega \sin b}{4\left(\frac{g^2 }{\omega}\cos^2b-J\right)},\\
        \cos^2b &=\frac{\omega}{ g^2}\left(J - \frac{\Omega}{4}\sqrt{1 - \frac{g^2}{\omega J}}\right).
    \end{aligned}
\end{equation}

As indicated by the derivations above, $m_{\mathrm{DK}}^x$ and the average photon number $n$ in Eq.~(\ref{mdx}) to be zero in the normal phase and nonzero in the superradiant phase, while  $m_{\mathrm{AF}}^z$ in Eq.~(\ref{mdnu}) is zero in the paramagnetic phase and nonzero in the antiferromagnetic phase. Specifically, in the AFS phase, all three quantities,  $m_{\mathrm{DK}}^x$, $n$, and  $m_{AF}^z$ are nonzero. This behavior is consistently observed and confirmed in all cases discussed below. 

In addition,  the stability of these ground-state phases should be examined by analyzing the Hessian matrix $\mathbf{H}_s = \nabla\nabla E_g$
\begin{equation}
    \mathbf{H}_s = \begin{pmatrix}
        H_{1} &H_d\\
        H_d &H_{2}
    \end{pmatrix},
\end{equation}
where $H_d = -\frac{2g^2}{\omega}\sin \theta_1 \sin \theta_2-J\cos \theta_1 \cos \theta_2$ and $H_{1(2)} = \frac{2g^2}{\omega}[(\sin \theta_{1(2)}+\cos\theta_{1(2)})\cos \theta_{2(1)}-2\sin^2 \theta_{1(2)}+1]$. The stability of the ground-state solutions requires the Hessian matrix to be positive definite, which implies all eigenvalues are positive. The stability conditions of the three ground-state phases are given as follows.

\textbf{(i)} PN phase:
\begin{equation}
    g < g_{c1}^{z,\mathrm{GS}} =\sqrt{\frac{\omega\left(\Omega-4J\right)}{4}},\;J<\frac{\Omega}{4}.
\end{equation}

\textbf{(ii)} AFN phase:
\begin{equation}
    g < g_{c2}^{z,\mathrm{GS}} = \sqrt{\frac{\omega\left(16J^2-\Omega^2\right)}{16J}},\; J>\frac{\Omega}{4}.
\end{equation}

\textbf{(iii)} PSR phase:
\begin{equation}
    16J_{\mathrm{eff}}^3-32J J_{\mathrm{eff}}^2 + J\Omega^2 >0,\;g >g_{c1}^{z,\mathrm{GS}}.
\end{equation}

\textbf{(iv)} AFS phase:
\begin{equation}
    16J_{\mathrm{eff}}^3-32J J_{\mathrm{eff}}^2 + J\Omega < 0,\;g > g_{c2}^{z,\mathrm{GS}}.
\end{equation}

\emph{Longitudinal DIM}-:
Considering the Hamiltonian \eqref{Hamiltonian1} for the longitudinal DIM without dissipation, we adopt a procedure analogous to that used for transverse DIM and obtain the stationary point equations as
\begin{equation}
    \begin{aligned}
        \sin a\left[2\left(\frac{g^2}{\omega}\cos^2 b -J\right)\cos a - \frac{\Omega}{2}\sin b\right] &=0,\\
        \cos b\left[2\left(\frac{g^2}{\omega}\cos^2 a - J\right)\sin b +\frac{\Omega}{2}\cos a\right] &= 0.
    \end{aligned}
\end{equation}

The solutions for minimizing ground-state energy can be classified as follows.

\textbf{(i)} PN phase ($\sin a = \cos b = 0$):
\begin{equation}
	 s_A^x = s_B^x = 0,\; s_A^z = s_B^z = -1,\;\alpha = 0.
\end{equation}

\textbf{(ii)} AFN phase ($\sin a \neq 0$ and $\cos b = 0$):
\begin{equation}
	 s_A^x = -s_B^x = \pm \sqrt{1-\frac{\Omega^2}{16J^2}},\; s_A^z = s_B^z = -\frac{\Omega}{4J},\;\alpha = 0.
\end{equation}

\textbf{(iii)} PS phase ($\sin a =0$ and $\cos b \neq0$):
\begin{equation}
    \begin{aligned}
        s_A^x &= s_B^x = \pm \sqrt{1-\frac{\Omega^2}{16J_{\mathrm{eff}}^2}} , \; s_A^z = s_B^z = \frac{\Omega}{4J_{\mathrm{eff}}},\\
        \alpha &= - \frac{g}{\omega}s_{A(B)}^x, 
    \end{aligned}
\end{equation}
where $J_{\mathrm{eff}} = J -   g^2/\omega$. 

\textbf{(iv)} AFS phase ($\sin a \neq 0$ and $\cos b \neq0$): 
\begin{equation}
    \begin{aligned}
        \cos a &= \frac{\Omega \sin b}{4\left(\frac{g^2}{\omega}\cos^2b-J\right)},\\
        \cos^2b &=\frac{\omega}{g^2}\left(J + \frac{\Omega}{4}\sqrt{\frac{g^2}{\omega J}-1}\right)
    \end{aligned}.
\end{equation}

Now, the stability of these ground state phases should be examined by analyzing the Hessian matrix, as given below.

\textbf{(i)} PN phase:  
\begin{equation}
	g < g_{c1}^{x,\mathrm{GS}} = \sqrt{\omega\left(J+\frac{\Omega}{4}\right)} ,\;J < \frac{\Omega}{4}.
\end{equation}

\textbf{(ii)} AFN phase:
\begin{equation}	           
    g<g_{c2}^{x,\mathrm{GS}}=\sqrt{\omega\frac{J\left(16J^2+\Omega^2\right)}{\Omega^2}}, 		J > \frac{\Omega}{4}.
\end{equation}

\textbf{(iii)} PS phase:
\begin{equation}
	g >g_{c1}^{x,\mathrm{GS}}, \; g>g_{c3}^{x,\mathrm{GS}}=\sqrt{\omega\left[J+\left(\frac{J\Omega^2}{16}\right)^{1/3}\right]}.
\end{equation}

This AFS solution is not a local minimum value and does not correspond to ground-state. Note that both the solutions of AFN and PS phases correspond to local minima of the ground-state energy in the interval $g_{c3}^{x,\mathrm{GS}} < g < g_{c2}^{x,\mathrm{GS}}$ when $J>\Omega/4$. It is therefore necessary to compare their energies to determine the true ground state. The ground-state energies for AFN and PS phases are given by
\begin{equation}
	\begin{cases}
		E_g^{\mathrm{AFN}} = -J - \frac{\Omega^2}{16J},\\
		E_g^\mathrm{PS} = J_{\mathrm{eff}} + \frac{\Omega^2}{16J_{\mathrm{eff}}}.
	\end{cases}
\end{equation}
A new critical coupling strength $g_{c0}^{x,\mathrm{GS}} = \sqrt{2\omega J}$ emerges: for $g < g_{c0}^{x,\mathrm{GS}}$, the AFN phase is the ground state of the system, whereas for $g > g_{c0}^{x,\mathrm{GS}}$, the PS phase becomes the ground state.

\section{\label{appendix steady-state solution} THE STEADY-STATE SOLUTIONS OF THE TRANSVERSE DIM AND LONGITUDINAL DIM}

In this Appendix, we present detailed derivations of the steady-state solutions for the transverse DIM \eqref{steady-state_z} and the longitudinal DIM \eqref{steady-state_x}.

\emph{Transverse DIM}-:
Starting from Eq.~\eqref{steady-state_z}, one can directly obtain the relation between the real part of the cavity field $\alpha_R$ and the spin as
\begin{equation}
    \alpha_R = -\frac{\omega g}{2\left(\omega^2 + \kappa^2\right)}\left(s_A^x + s_B^x\right).  
    \label{alpha_R}
\end{equation}
Substituting Eq.~\eqref{alpha_R} into Eq.~\eqref{steady-state_z}, we derive the steady-state equation involving only the spin degrees of freedom,
\begin{equation}
    \left(\Omega + 4Js_{B(A)}^z\right)s_{A(B)}^x + \frac{2\omega g^2}{\omega^2+\kappa^2}\left(s_A^x + s_B^x\right)s_{A(B)}^z = 0.
\end{equation}

Due to the antiferromagnetic interaction, we parameterize the spin components as $s_{A(B)}^x = \cos \theta_{1(2)}$ and $s_{A(B)}^z=\pm \sin\theta_{1(2)}$, which leads to the following two steady-state equations:
\begin{equation}
    \begin{aligned}
        \left(\Omega - 4J \sin\theta_2\right)\cos\theta_1 + \frac{2\omega g^2}{\omega^2 + \kappa^2}\left(\cos \theta_1 + \cos \theta_2\right) \sin \theta_1 &=0,\\
        \left(\Omega + 4J \sin\theta_1\right)\cos\theta_2 - \frac{2\omega g^2}{\omega^2 + \kappa^2}\left(\cos \theta_1 + \cos \theta_2\right) \sin \theta_2 &=0.
    \end{aligned}
    \label{steady-state_theta_z}
\end{equation}

We introduce the symmetric and antisymmetric combinations $a=(\theta_1 + \theta_2)/2$ and $b=(\theta_1-\theta_2)/2$ like Appendix.~\ref{appendix-ground} to simplify Eq.~\eqref{steady-state_theta_z} as
\begin{equation}
    \begin{aligned}
        0 &=\frac{4\omega g^2}{\omega^2 +\kappa^2}\sin(a+b) \cos a \cos b  \\
        &- 2J\left(\sin2a - \sin2b\right) 
        + \Omega \cos(a+b),\\
        0 &=\frac{4\omega g^2}{\omega^2 +\kappa^2}\sin(a-b) \cos a  \cos b\\
        &- 2J\left(\sin2a + \sin2b\right)
        - \Omega \cos(a-b).
    \end{aligned}
    \label{steady_ab_z}
\end{equation}

By rearranging Eq.~\eqref{steady_ab_z}, we finally obtain
\begin{equation}
    \begin{aligned}
        \sin a\left[4\left(\frac{\omega g^2}{\omega^2+\kappa^2}\cos^2b-J\right)\cos a - \Omega\sin b\right] &=0,\\
        \cos b\left[4\left(\frac{\omega g^2}{\omega^2+\kappa^2}\cos^2a +J\right)\sin b + \Omega\cos a\right] &=0.
    \end{aligned}
\end{equation}

The solutions can be classified into four distinct cases.

\textbf{(i)} PN phase ($\sin a = \cos b =0$):
\begin{equation}
	 s_A^x = s_B^x = 0,\; s_A^z = s_B^z = -1,\;\alpha = 0.
\end{equation}

\textbf{(ii)} AFN phase ($\sin b = \cos a=0$): 
\begin{equation}
    s_A^x = s_B^x=0,\;s_A^z=-s_B^z= \pm1,\;\alpha=0.
\end{equation}

\textbf{(iii)} PS phase ($\sin a=0$ and $\cos b \neq 0$): 
\begin{equation}
    \begin{aligned}
        s_A^x&=s_B^x=\pm\sqrt{1-s_{A(B)}^{z2}},\;s_A^z = s_B^z = - \frac{\Omega}{4J_{\mathrm{eff}}},\\
        \alpha&=-\frac{g}{\omega}s_{A(B)}^x,
    \end{aligned}
\end{equation}
where $J_{\mathrm{eff}} = J+ \omega g^2/(\omega^2+\kappa^2)$.

\textbf{(iv)} AFS phase ($\sin a \neq0$ and $\cos b \neq0$): 
\begin{equation}
    \begin{aligned}
        \cos a &= \frac{\Omega \sin b}{4\left(\frac{\omega g^2 \cos^2b}{\omega^2+\kappa^2}-J\right)},\\
        \cos^2b &=\frac{\omega^2 + \kappa^2}{\omega g^2}\left(J - \frac{\Omega}{4}\sqrt{1 - \frac{\omega g^2}{J\left(\omega^2 + \kappa^2 \right)}}\right).
    \end{aligned}
\end{equation}

\emph{Longitudinal DIM}-:
Starting from Eq.~\eqref{steady-state_x}, we can readily obtain a steady-state equation involving only the spin degrees of freedom,
\begin{equation}
    \Omega s_{A(B)}^x - 4\left[Js_{B(A)}^x - \frac{\omega g^2}{2\left(\omega^2 + \kappa^2\right)}\left(s_A^x + s_B^x\right)\right]s_{A(B)}^z=0.
\end{equation}

Following the same procedure as for the transverse DIM, we also obtain the steady-state equation for $a$ and $b$:

\begin{equation}
    \begin{aligned}
        \sin a\left[4\left(\frac{\omega g^2}{\omega^2 + \kappa^2} \cos^2b-J\right)\cos a - \Omega \sin b\right] &= 0,\\
        \cos b \left[4\left(\frac{\omega g^2}{\omega^2 + \kappa^2} \cos^2 a-J\right)\sin b + \Omega\cos a\right] &=0
    \end{aligned}
\end{equation}

The solutions can be classified into four distinct cases.

\textbf{(i)} PN phase ($\sin a = 0$ and $\cos b = 0$):
\begin{equation}
	 s_A^x = s_B^x = 0,\; s_A^z = s_B^z = -1,\;\alpha = 0.
\end{equation}

\textbf{(ii)} AFN phase ($\sin a \neq 0$ and $\cos b = 0$):
\begin{equation}
	 s_A^x = -s_B^x = \pm \sqrt{1-\frac{\Omega^2}{16J^2}},\; s_A^z = s_B^z = -\frac{\Omega}{4J},\;\alpha = 0.
\end{equation}

\textbf{(iii)} PS phase ($\sin a =0$ and $\cos b \neq0$):
\begin{equation}
    \begin{aligned}
        s_A^x &= s_B^x = \pm \sqrt{1-\frac{\Omega^2}{16J_{\mathrm{eff}}^2}} , \; s_A^z = s_B^z = \frac{\Omega}{4J_{\mathrm{eff}}},\\
        \alpha &= - \frac{g}{\omega}s_{A(B)}^x, 
    \end{aligned}
\end{equation}
where $J_{\mathrm{eff}} = J -   \omega g^2/(\omega^2 + \kappa^2)$. 

\textbf{(iv)} AFS phase: $\sin a \neq 0$
and $\cos b \neq0$
\begin{equation}
    \begin{aligned}
        \cos a &= \frac{\Omega \sin b}{4\left(\frac{\omega g^2}{\omega^2 + \kappa^2}\cos^2b-J\right)},\\
        \cos^2b &=\frac{\omega^2 + \kappa^2}{\omega g^2}\left(J + \frac{\Omega}{4}\sqrt{\frac{\omega g^2}{J\left(\omega^2 + \kappa^2\right)}-1}\right)
    \end{aligned}.
\end{equation}

However, this AFS solution is dynamically unstable and therefore does not correspond to a physical steady state.

\section{\label{appendix stability analysis} STABILITY ANALYSIS OF THE DISSIPATIVE TRANSVERSE DIM AND LONGITUDINAL DIM}

\emph{Transverse DIM}-:
To examine the dynamical stability of the mean-field solutions, small fluctuations around the mean-field steady states are introduced as $\alpha \rightarrow \alpha + \delta\alpha$, and $s_\mu^\nu \rightarrow s_\mu^\nu + \delta s_\mu^\nu$. Retaining only terms linear in these fluctuations yields a set of linearized equations of motion governing the dynamics of small perturbations. The fluctuation vector is defined as $\delta \mathbf{X} = (\delta\alpha_R, \delta\alpha_I, \delta s_A^x, \delta s_A^y, \delta s_A^z, \delta s_B^x, \delta s_B^y, \delta s_B^z)^\top$, and its time evolution is governed by $\delta \dot{\mathbf{X}} = \mathbf{M} \delta \mathbf{X}$, where $\mathbf{M}$ is dynamical matrix. Using the spin-length constraint, the fluctuations satisfy $\delta s_\nu^z = -\frac{1}{s_\mu^z}\left(s_\mu^x\,\delta s_\mu^x + s_\mu^y\,\delta s_\mu^y\right)=-\frac{s_\mu^x}{s_\mu^z}\delta s_\mu^x$, which allows us to eliminate $\delta s_\mu^z$. The reduced fluctuation vector is therefore $\delta\mathbf{X}_\mathrm{re} = (\delta\alpha_R, \delta\alpha_I, \delta s_A^x, \delta s_A^y, \delta s_B^x, \delta s_B^y)^\top$, and the reduced dynamical equation reads $\delta \dot{\mathbf{X}}_{\mathrm{re}}= \mathbf{M_{\mathrm{re}}} \delta \mathbf{X}_\mathrm{re}$ with
\begin{equation}
	\mathbf{M}_{\mathrm{re}} = 
	\begin{pmatrix}
		-\kappa & \omega & 0 & 0 & 0 & 0 \\
		-\omega & -\kappa & -g/2 & 0 & -g/2 & 0 \\
		0 & 0 & 0 & -\Omega_A & 0 & 0 \\
		-4g s_{A}^z & 0 & \chi_A & 0 & \eta_A & 0 \\
		0 & 0 & 0 & 0 & 0 & -\Omega_B  \\
		-4g s_B^z & 0 &\eta_B & 0 &\chi_B  & 0
	\end{pmatrix},
\end{equation}
where $\Omega_{A(B)}=\Omega + 4Js_{B(A)}^z$, $\eta_{A(B)}=-4J{s_A^xs_B^x}/{s_{B(A)}^z}$, and $\chi_{A(B)} = \Omega_{A(B)}+4g\alpha_Rs_{A(B)}^x/s_{A(B)}^z$. For PN and PS phases, $\Omega_{A(B)}=\Omega_{\mathrm{eff}}$ $\chi_{A(B)} = \chi$, $\eta_{A(B)} = \eta$, and $s_A^z = s_B^z = s^z$. Under these conditions, the stability matrix can be rewritten as
\begin{equation}
	\mathbf{M}_{\mathrm{re}} = 
	\begin{pmatrix}
		\mathcal{A} & \mathcal{B} & \mathcal{B} \\
		\mathcal{C} & \mathcal{D} & \mathcal{E} \\
		\mathcal{C} & \mathcal{E} & \mathcal{D}
	\end{pmatrix},
\end{equation}
with
\begin{equation}
    \begin{aligned}
    &\mathcal{A} = 
	\begin{pmatrix}
		-\kappa & \omega \\
		-\omega & -\kappa
	\end{pmatrix}, \;
	\mathcal{B} = 
	\begin{pmatrix}
		0 & 0 \\
		-g/2 & 0
	\end{pmatrix}, \;
	\mathcal{C} = 
	\begin{pmatrix}
		0 & 0 \\
		-4g s^z & 0
	\end{pmatrix}, \;\\
	&\mathcal{D} = 
	\begin{pmatrix}
		0 & -\Omega_{\mathrm{eff}} \\
		\chi & 0
	\end{pmatrix}, \;
	\mathcal{E} = 
	\begin{pmatrix}
		0 & 0 \\
		\eta & 0
	\end{pmatrix}.
    \end{aligned}
\end{equation}
Consequently, the fluctuation vector can be decomposed as $\delta \mathbf{X}_{\mathrm{re}} = \left(\delta \mathbf{X}_0, \delta \mathbf{X}_A, \delta \mathbf{X}_B \right)^\top$, where $\delta \mathbf{X}_0 = (\delta \alpha_R, \delta \alpha_I)^\top$ and $\delta \mathbf{X}_{A(B)} = (\delta s_{A(B)}^x, \delta s_{A(B)}^y)^\top$, correspond to cavity-field fluctuations and spin fluctuations on the $A$ ($B$) sublattice, respectively. 

To further simplify the analysis, we perform a unitary transformation $\mathbf{M}_{\mathrm{re}}^\prime = \mathbf{P} \mathbf{M}_{\mathrm{re}} \mathbf{P}^{-1}$ with 
\begin{equation}
	\mathbf{P} = 
	\begin{pmatrix}
		\mathbf{I}_2 & 0 & 0 \\
		0 & \mathbf{I}_2/\sqrt{2} & \mathbf{I}_2/\sqrt{2} \\
		0 & \mathbf{I}_2/\sqrt{2} & -\mathbf{I}_2/\sqrt{2}
	\end{pmatrix},
\end{equation}
where $\mathbf{I}_2$ denotes the $2 \times 2$ identity matrix. Accordingly, the fluctuation basis transforms to $\delta \mathbf{X}_{\mathrm{re}}^\prime = P \delta \mathbf{X}_{\mathrm{re}} = \left(\delta \mathbf{X}_0,\, \frac{1}{\sqrt{2}}\left(\delta \mathbf{X}_A + \delta \mathbf{X}_B\right),\,\frac{1}{\sqrt{2}}\left(\delta \mathbf{X}_A - \delta \mathbf{X}_B\right)\right)$.
On this basis, the dynamical matrix becomes
\begin{equation}
	\mathbf{M}_{\mathrm{re}}^{\prime} =
	\begin{pmatrix}
		\mathcal{A} & \sqrt{2}\mathcal{B} & 0 \\
		\sqrt{2} \mathcal{C} & \mathcal{D} + \mathcal{E} & 0 \\
		0 & 0 & \mathcal{D} - \mathcal{E}
	\end{pmatrix}
	=
	\begin{pmatrix}
		\mathbf{N}_1 & 0 \\
		0 & \mathbf{N}_2
	\end{pmatrix},
\end{equation}
where 
\begin{equation}
    \begin{aligned}
    	\mathbf{N}_1 &= 
	\begin{pmatrix}
		-\kappa &\omega &0 &0\\
		-\omega &-\kappa &-g/\sqrt{2} &0\\
		0 &0 &0 &-\Omega_{\mathrm{eff}}\\
		-4\sqrt{2}gs^z &0 &\chi + \eta &0
	\end{pmatrix},\\
	\mathbf{N}_2 &= 
	\begin{pmatrix}
		0 &-\Omega_{\mathrm{eff}}\\
		\chi-\eta &0
	\end{pmatrix}.    
    \end{aligned}
\end{equation}
which is explicitly block diagonal. As a result, the dynamical stability of the two phases can be determined by analyzing the eigenvalue spectra of the two independent blocks $\mathbf{N}_1$ and $\mathbf{N}_2$. The eigenvalues of $\mathbf{N}_2$ can be readily obtained as $\lambda_{1,\pm} = \pm i\sqrt{\Omega_{\mathrm{eff}}\left(\chi -\eta\right)}$ and the corresponding stability condition is given by $\Omega_{\mathrm{eff}}\left(\chi-\eta\right) > 0$. The eigenvalues of $\mathbf{N}_1$ are determined by the characteristic equation
\begin{equation}
	\lambda^4 + a_1 \lambda^3 + a_2 \lambda^2 + a_3\lambda + a_4 = 0 .
\end{equation}
Here, $a_1 = 2\kappa$, $a_2 = (\omega ^{2}+\kappa ^{2} + \xi_1$, $a_3 = 2\kappa \xi_1$, and $a_4 = \xi_1 (\omega ^{2}+\kappa ^{2}) + \xi_2$, where $\xi_1 = \Omega_{\mathrm{eff}} \left(\chi +\eta\right)$ and $\xi_2 = 4\Omega_{\mathrm{eff}} \omega g^2 s^z$. The stability of the system can then be analyzed using the Routh–Hurwitz criterion. The stability conditions require
$a_n > 0$ $(n = 1, 2, 3, 4)$,
$\Delta_2 = a_1 a_2 - a_3 > 0$,
and $\Delta_3 = a_1 a_2 a_3 - a_3^2 - a_1^2 a_4 > 0$, i.e.,
\begin{equation}
	\begin{cases}
			&\xi_1 = \Omega_{\mathrm{eff}} \left(\chi +\eta\right) >0,\\
			&\xi_2 = 4\Omega_{\mathrm{eff}} \omega g^2 s^z <0,\\
			&\xi_1 (\omega ^{2}+\kappa ^{2}) + \xi_2 > 0.
	\end{cases}
\end{equation}
Finally, we obtain the stability conditions for PN and PS phases as follows.

\textbf{(i)} PN phase
\begin{equation}
    g<g_{c1}^z =\sqrt{\frac{\omega\left(\Omega-4J\right)\left(\omega^2+\kappa^2\right)}{4\omega}},\;J<\frac{\Omega}{4}.
\end{equation}

\textbf{(iii)} PSR phase
\begin{equation}
    16J_{\mathrm{eff}}^3-32J J_{\mathrm{eff}}^2 + J\Omega^2 >0,\;g>g_{c1}^z
\end{equation}
Here, $J_{\mathrm{eff}} = J+ \omega g^2/(\omega^2+\kappa^2)$.

We next analyze the stability conditions for the AFN and AFS phases. 
The reduced dynamical matrix for the AFN phase is given by
\begin{equation}
	\mathbf{M}_{\mathrm{re}}^{\mathrm{AFN}} = 
	\begin{pmatrix}
		-\kappa & \omega & 0 & 0 & 0 & 0 \\
		-\omega & -\kappa & -g/2 & 0 & -g/2 & 0 \\
		0 & 0 & 0 & -\Omega_A & 0 & 0 \\
		-4g & 0 & -\Omega_A & 0 & 0 & 0 \\
		0 & 0 & 0 & 0 & 0 & -\Omega_B \\
		4g & 0 & 0 & 0 & \Omega_B & 0
	\end{pmatrix},
\end{equation}
where $\Omega_{A(B)} = \Omega \mp 4J$. We assume fluctuations of the form $\delta\mathbf{X}_{\mathrm{re}} = \mathbf{v} e^{\lambda t}$ and substitute this ansatz into the linearized dynamical equation
$\delta\dot{\mathbf{X}}_{\mathrm{re}}^{\mathrm{AFN}} = \mathbf{M}_{\mathrm{re}}^{\mathrm{AFN}} \delta\mathbf{X}_{\mathrm{re}}$.
This leads to the characteristic equation
\begin{equation}
    \left[(\lambda + \kappa)^2 + \omega^2\right]
    + 2\omega g^2
    \left(
    \frac{\Omega_A}{\lambda^2 + \Omega_A^2}
    - \frac{\Omega_B}{\lambda^2 + \Omega_B^2}
    \right)
    = 0.
\end{equation}

To determine the instability boundary, we set $\lambda = 0$, yielding
\begin{equation}
    g_{c2}^z
    =
    \sqrt{
    \frac{(\omega^2 + \kappa^2)(16J^2 - \Omega^2)}
    {16\omega J}
    } .
\end{equation}

Finally, the stability conditions for the AFN and AFS phases are obtained accordingly.

\textbf{(ii)} AFN phase:
\begin{equation}
    g<g_{c2}^z,\; J>\frac{\Omega}{4}.
\end{equation}

\textbf{(iv)} AFS phase:
\begin{equation}
    16J_{\mathrm{eff}}^3-32J J_{\mathrm{eff}}^2 + J\Omega < 0,\;g>g_{c2}^z.
\end{equation}

\emph{Longitudinal DIM}-:
 The reduced dynamical matrix $\mathbf{M}_{\mathrm{re}}$ for longitudinal DIM is given by

\begin{equation}
	\mathbf{M}_{\mathrm{re}} = 
	\begin{pmatrix}
		-\kappa & \omega & 0 & 0 & 0 & 0 \\
		-\omega & -\kappa & -g/2 & 0 & -g/2 & 0 \\
		0 & 0 & 0 & -\Omega & 0 & 0 \\
		-4g s_{A}^z & 0 & \chi_A & 0 & -4J s_A^z & 0 \\
		0 & 0 & 0 & 0 & 0 & -\Omega \\
		-4g s_B^z & 0 & -4J s_B^z & 0 & \chi_B & 0
	\end{pmatrix},
\end{equation}
where $\chi_{A(B)}=\Omega+4(g\alpha_R+Js_{B(A)})s_{A(B)}^x/s_{A(B)}^z$. Note that $\chi_{A(B)} = \chi$ and $s_A^z = s_B^z = s^z$ hold in the PN, AFN, and PS phases. Similarly, the dynamical stability is determined by the independent matrix
\begin{equation}
    \begin{aligned}
    	\mathbf{N}_1 &= 
	\begin{pmatrix}
		-\kappa &\omega &0 &0\\
		-\omega &-\kappa &-g/\sqrt{2} &0\\
		0 &0 &0 &-\Omega\\
		-4\sqrt{2}gs^z &0 &\chi-4Js^z &0
	\end{pmatrix},\\
	\mathbf{N}_2 &= 
	\begin{pmatrix}
		0 &-\Omega\\
		\chi + 4Js^z &0
	\end{pmatrix}.    
    \end{aligned}
\end{equation}

The eigenvalues of $\mathbf{N}_2$ can be readily obtained as $\lambda_{1,\pm} = \pm i\sqrt{\Omega\left(\chi + 4Js^z\right)}$ and the corresponding stability condition is given by $\chi + 4Js^z > 0$. The eigenvalues of $\mathbf{N}_1$ are determined by the characteristic equation
\begin{equation}
	\lambda^4 + a_1 \lambda^3 + a_2 \lambda^2 + a_3\lambda + a_4 = 0 .
\end{equation}
Here, $a_1 = 2\kappa$, $a_2 = \kappa^2 + \omega^2 + \xi_1$, $a_3 = 2\kappa \xi_1$, and $a_4 = \xi_1 \left(\kappa^2 + \omega^2\right) + \xi_2$, where $\xi_1 = \Omega \left(\chi - 4Js^z\right)$ and $\xi_2 = 4\Omega \omega g^2 s^z$. According to the Routh-Hurwitz stability criterion, the AFS solution is dynamically unstable. 
The stability conditions for the PN, AFN, and PS phases are thus obtained as follows.

\textbf{(i)} PN phase:
\begin{equation}
	\begin{aligned}
		g < g_{c1}^x=\sqrt{\frac{(\Omega + 4J)(\omega ^{2}+\kappa ^{2})}{4\omega}}, 
		\; J < \frac{\Omega}{4}.
	\end{aligned}
\end{equation}

\textbf{(ii)} AFN phase:
\begin{equation}
	\begin{aligned}
		g < g_{c2}^x =\sqrt{\frac{J(16J^2 + \Omega^2)(\omega ^{2}+\kappa ^{2})}{\omega \Omega^2}},\; J > \frac{\Omega}{4}.
	\end{aligned}
\end{equation}

\textbf{(iii)} PS phase:
\begin{equation}
	g >g_{c1}^x,\;g>g_{c3}^x=\sqrt{\dfrac{\omega^2 + \kappa^2}{\omega}\left[J + \left(\dfrac{J\Omega^2}{16}\right)^{1/3}\right]}.
\end{equation}

\bibliographystyle{MSP}
\bibliography{refs}
\end{document}